\documentclass[apjl]{emulateapj}
\usepackage{natbib}
\usepackage{hyperref}
\hypersetup{
    colorlinks=true,
    linkcolor=blue,
    citecolor=blue,
    filecolor=blue,
    urlcolor=blue
}
\usepackage{amsmath,amssymb}
\usepackage{mathptmx}
\usepackage{mathtools}
\DeclareMathAlphabet{\pazocal}{OMS}{zplm}{m}{n}

\usepackage{graphicx}
\usepackage{tabularx}
\defcitealias{Leike2020a}{L20}
\defcitealias{Ungerechts1987}{UT87}
\begin{document}

\title{
The Per-Tau Shell: \\
A Giant Star-Forming Spherical Shell Revealed by 3D Dust Observations
}

\author{Shmuel Bialy\altaffilmark{1}$^{\star}$,  Catherine Zucker\altaffilmark{1}, Alyssa Goodman\altaffilmark{1,2},Michael M. Foley\altaffilmark{1}, Jo\~{a}o Alves\altaffilmark{2,3}, 
Vadim A. Semenov\altaffilmark{1},
Robert Benjamin\altaffilmark{4},
Reimar Leike\altaffilmark{5,6}, Torsten En${\ss}$lin\altaffilmark{5,6}}
 
\altaffiltext{1}{Center for Astrophysics $\vert$ Harvard \& Smithsonian, 60 Garden st., Cambridge, MA, 02138
 }
 \altaffiltext{2}{Radcliffe Institute for Advanced Study, Harvard University, 10 Garden Street, Cambridge, MA 02138, USA}
 \altaffiltext{3}{University of Vienna, Department of Astrophysics, Türkenschanzstrasse 17, 1180 Wien, Austria}
 \altaffiltext{4}{Department of Physics, University of Wisconsin–Whitewater, 800 West
Main Street, Whitewater, WI 53190}
  \altaffiltext{5}{Max Planck Institute for Astrophysics, Karl-Schwarzschildstra{\ss}e 1,85748 Garching, Germany}
    \altaffiltext{6}{Ludwig-Maximilians-Universität, Geschwister-SchollPlatz1, 80539 Munich, Germany}


  \email{$^\star$sbialy@cfa.harvard.edu}
\slugcomment{
}

\begin{abstract}

A major question in the field of star formation is how molecular clouds form out of the diffuse Interstellar Medium (ISM).
Recent advances in 3D dust mapping are revolutionizing our view of the structure of the ISM.
Using the highest-resolution 3D dust map to date, we explore the structure of a nearby star-forming region, which includes the well-known Perseus and Taurus molecular clouds.
We reveal an extended near-spherical shell, 156 pc in diameter, hereafter the ``Per-Tau Shell", in which the Perseus and Taurus clouds are embedded. We also find a large ring structure at the location of Taurus, hereafter, the ``Tau Ring".
We discuss a formation scenario for the Per-Tau Shell, in which previous stellar and supernova (SN) feedback events formed a large expanding shell, where the swept-up ISM has condensed to form both the shell and the Perseus and Taurus molecular clouds within it.
We present auxiliary observations of HI, H$\alpha$, $^{26}$Al, and X-rays that further support this scenario, and estimate Per-Tau Shell's age to be $\approx 6-22$ Myrs.
The Per-Tau shell offers the first three-dimensional observational view of a phenomenon long-hypothesized theoretically,
molecular cloud formation and star formation triggered by previous stellar and SN feedback.
\end{abstract}

\keywords{Interstellar medium (847) -- Molecular clouds (1072) --  Solar neighborhood (1509) -- Stellar feedback (1602) -- Superbubbles (1656) -- Astronomy data visualization (1968)}

\section{Introduction}
\label{sec: intro}
The conversion of diffuse gas into a dense-cold phase is the first step and a potential bottleneck for star formation.
The balance between heating and cooling processes in the ISM results in a multiphase gas, consisting of the cold/warm neutral media \citep[CNM/WNM;][]{Field1969, Wolfire2003, Seifried2010, Saury2014, Bialy2019}.
The CNM, being colder and denser, is the phase susceptible to molecule formation, gravitational collapse and star formation.
Numerical simulations have shown that the WNM-to-CNM conversion takes place in converging flows where WNM is
shock compressed and then radiatively cools, forming CNM \citep[][]{Hennebelle1999, Koyama2000, Koyama2002, Audit2005, Vazquez-Semadeni2006, Heitsch2006, Inoue2008, Inoue2009, Hennebelle2019}.
These flows naturally occur in expanding superbubbles (SB) powered by SN and winds from massive stars, introducing a positive feedback-loop for star formation:
[star formation] $\rightarrow$ [expanding shell] $\rightarrow$ [gas compression] $\rightarrow$ [gravitational collapse] $\rightarrow$ [star formation]
\citep[e.g.,][]{Hartmann2002, Hosokawa2006, Ntormousi2011, Palous2014a}.

\begin{figure*}
    \centering
    \includegraphics[width=1.0\textwidth]{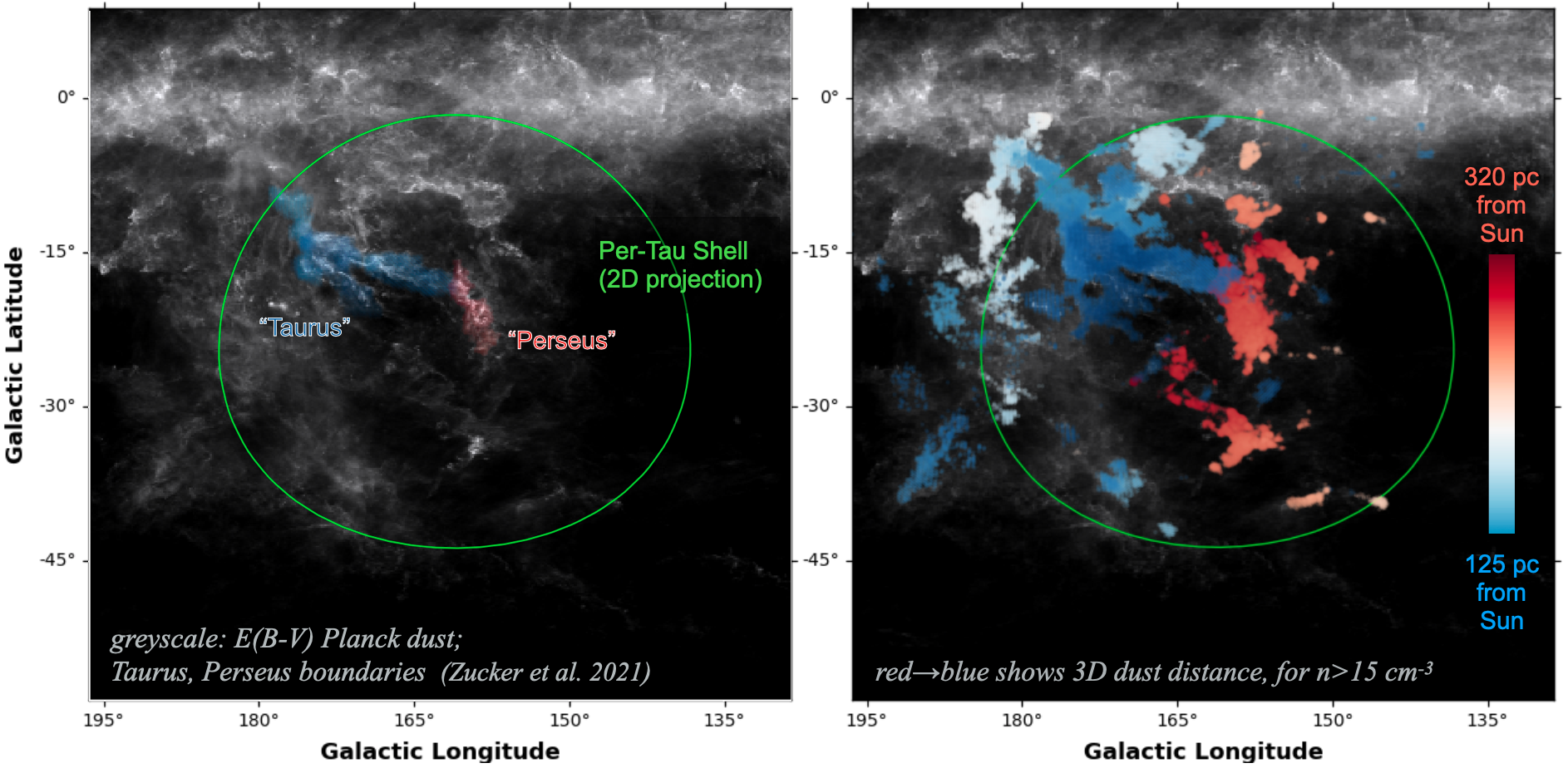}
    \caption{Finding Chart. 
    {\bf Left}: The Perseus-Taurus region as seen with Planck E(B-V) dust \citep[][]{Abergel2014}. 
    The Taurus and Perseus molecular clouds are indicated  \citep[as defined in][]{Zucker2021_topology}.
    {\bf Right}: 2D Projection of 3D density structures (densities $n>15$ cm$^{-3}$) based on \citetalias{Leike2020a}'s 3D dust map.
    While on the plane of the sky (as well as in velocity space; see Fig.~\ref{fig: 2D sky}a), the Perseus and Taurus clouds seem to connect,
    in practice they are separated by $\approx 150$ pc along the LOS. 
    As we discuss in {\it this} paper, the two clouds are still related, as they both lie on two opposing ends of a large 3D shell, the ``Per-Tau Shell." 
    }
    \label{fig:finding chart}
\end{figure*}

The formation of dense gas and stars in expanding shells has been observed in CO and HI spectral \textit{position-position-velocity} (PPV) cubes and in 2D dust extinction maps
\citep[e.g.][see reviews by \citealt{Elmegreen2011,Dawson2013}]{Su2009, Lee2009, Dawson2011, Dawson2015, Mackey2017}.
However, a major limitation with 2D extinction maps and PPV cubes is that they do not include information on the distance along the line-of-sight (LOS), and the data are limited to the 2D ``plane-of-the sky."
This results in blending of structures along the LOS
and a limited access to the real 3D structure of gas in the ISM.
This is especially a problem for structures that are only marginally denser than the surrounding ISM, 
such as old SN remnants, which may be major drivers of WNM-CNM conversion, molecule formation and star formation triggering \citep[see Fig.~2 in][]{Inutsuka2015}.

Since the advent of \textit{Gaia} \citep{Brown2018}, rapid progress has been made in the field of 3D dust mapping, which leverages stellar distance and extinction estimates made by modeling photometry and astrometry to chart the 3D distribution of dust \citep{Lallement2019, Green2019, Chen2019, Leike2019}.
The basic idea is intuitive: 
starlight extinction measures the dust column density whereas a parallax measurement constrains the star's distance. Synthesizing column densities and distances towards a large number of stars provides information on the dust 3D density.
Recently, \citet[][herefter \citetalias{Leike2020a}]{Leike2020a} extended this framework, combining metric Gaussian variational inference with Gaussian Processes
to construct a 3D dust map of the solar neighborhood's ISM at 1 pc resolution.

In this paper, we use the \citetalias{Leike2020a} data to study the 3D density structure of the ISM in the Taurus-Perseus star-forming region (see Fig.~\ref{fig:finding chart}). 
This region is amongst the most thoroughly observed and studied star-forming regions in the Solar Neighborhood.
When large-scale mapping of molecular gas first became feasible, \citet[][hereafter \citetalias{Ungerechts1987}]{Ungerechts1987} mapped the Perseus-Taurus region in $^{12}$CO, and included a section in their paper entitled “Are the Taurus and IC 348 (a prominent part of Perseus) Clouds Connected?.” 
In velocity space, the CO data show a smooth connecting bridge between Taurus and Perseus, suggesting that a physically continuous bridge of gas connecting the clouds is possible. 
But, as \citetalias{Ungerechts1987} wisely suggested, velocity connection is not enough, and the plane-of-the-Sky “connection” of Taurus and Perseus could be a chance superposition.

We show that the connection between Taurus and Perseus is neither one of a physical bridge of gas or a “chance superposition.” Instead, Taurus and Perseus mark regions of compressed gas on opposite sides of an extended 3D SB, blown by SNe over the past $\sim 10$ Myr.

\section{Observational Data}
\label{sec: data}
We utilize the 3D dust extinction map recently published by \citetalias{Leike2020a}.
The data cover a region (740 pc)$^2$ wide in $xy$, and 570 pc tall in the $z$ direction, where $xyz$ are the Heliocentric Cartesian Galactic coordinates with the Sun at the origin.
For each voxel in the data cube, \citetalias{Leike2020a} provide the opacity density per parsec: 
$s_x \equiv (\Delta\tau_{\rm G})/(\Delta L/{\rm pc})$,
where $\tau_{\rm G}$ is the dust opacity in the \textit{Gaia} \textit{G} band, and 
$\Delta\tau_{\rm G}/\Delta L$ is the difference in the \textit{Gaia} \textit{G} band dust opacity per unit length.
The opacity density is proportional to the gas number density.
Assuming a standard extinction curve, 
$A_{\rm G}/N = 4\times 10^{-22}$ mag cm$^2$ \citep{Draine2011}, where $A_{\rm G}$ is \textit{Gaia's} \textit{G}-band extinction and $N$ is the hydrogen nuclei column density, we obtain
\begin{equation}
\label{eq: n-tau}
    n = 880 \  s_x \ {\rm cm^{-3}} \ .
\end{equation}
Here $n$ is the number density 
of hydrogen nuclei, including both atomic and molecular phases.
Applying this conversion factor to \citetalias{Leike2020a}'s 3D dust data, we get $n$ as a function of the 3D position, $(x,y,z)$ throughout the cube. In this paper we focus on a subcube of the full map, spanning $x=[-340, -40]$ pc, $y=[-83, 217]$ pc, $z=[-238,62]$ pc and centered on the mid-distance of Perseus and Taurus.

In addition to the 3D data, we also make use of various 2D and PPV observations.
We use Planck's $E(B-V)$ dust \citep{Abergel2014} and $^{12}$CO from \citet{Dame2001a} to explore the link between the 3D dust map and 2D observations. 
To trace past SN activity, we explore HI observations from HI4PI \citep{Bekhti2016}, 
H$\alpha$ observations from \cite{Finkbeiner2003}, 
$^{26}$Al observations from COMPTEL \citep{Diehl1995}, and X-ray observations from ROSAT \citep{Snowden1997} and from eROSITA's public release image (see Appendix \ref{app: Xrays}).

All this observational data, as well as our 3D models (\S \ref{sec: results}), are publicly available on Harvard's DataVerse\footnote{
\href{https://doi.org/10.7910/DVN/6ODS8M}{https://doi.org/10.7910/DVN/6ODS8M}.
Also included is the original \textit{glue} session, which can be used to reproduce the figures as well as for further exploration.}.

\begin{figure*}
    \centering
    \includegraphics[width=1\textwidth]{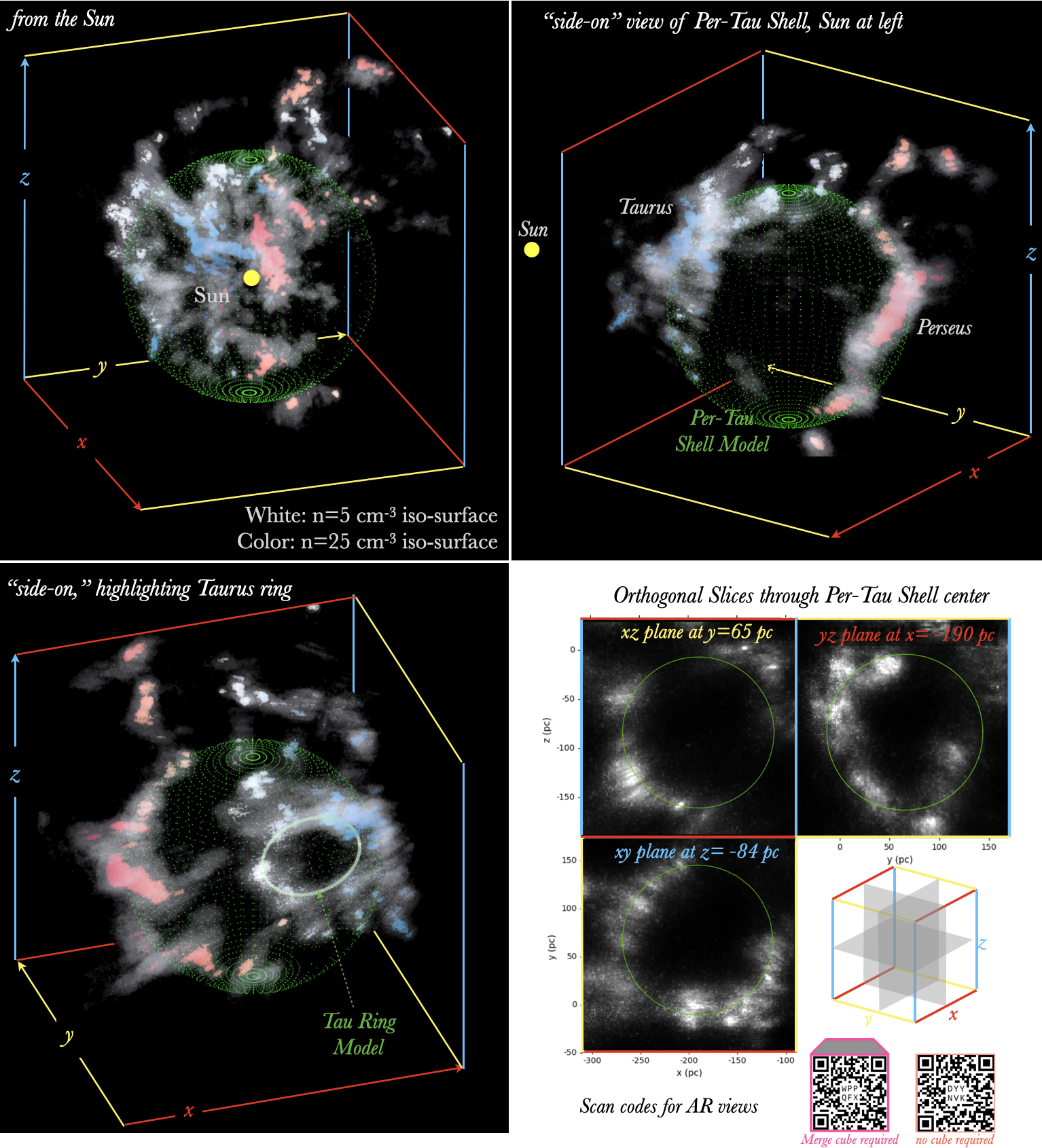}
    \caption{
    3D
    views of the Per-Tau shell (for an interactive version of this figure click \href{https://faun.rc.fas.harvard.edu/czucker/Paper\_Figures/sbialy/pertau\_superbubble.html}{here}
    \footnote{\label{footnote: interactive figure} URL address for the interactive figure: \href{https://faun.rc.fas.harvard.edu/czucker/Paper\_Figures/sbialy/pertau\_superbubble.html}{https://faun.rc.fas.harvard.edu/czucker/Paper\_Figures/sbialy/pertau\_superbubble.html}.} \footnote{\label{footnote: AR figure} For Augmented Reality (AR) experience, 
    scan the QR codes with a mobile device (CoSpaces App required): Left code for viewing with a \href{https://mergeedu.com/cube}{Merge Cube}, right code for a tabletop AR. 
    If you are already reading this article with a mobile device, click \href{https://edu.cospaces.io/WPP-QFX}{here} (Merge Cube), or \href{https://edu.cospaces.io/DYY-NVK}{here} (tabletop AR).
    }. See Fig.~\ref{fig: 3D plus} for more static visualizations).
    Plotted are density iso-surfaces at levels $n=5$ cm$^{-3}$  (grey) and $n=25$ cm$^{-3}$ (color), overlaid with our spherical shell model,radius $R_s=78$ pc,  distance from the sun $d=218$ pc.    
    The $n=25$ cm$^{-3}$ surfaces are colored by distance from the sun (blue-to-red).
    {\bf Upper-left}: View from the sun (compare with Fig.~\ref{fig:finding chart}). {\bf Upper-right}: A side view of the region. Perseus and Taurus and their diffuse envelopes are arranged on two opposing sides of the  Per-Tau shell.
     {\bf Lower-left}: Another side view emphasizing the Tau Ring. The ellipse is the Tau Ring model (Appendix \ref{app: ring}).
    {\bf Lower-right}: 2D density slices along the $xy$, $xz$, $yz$ planes. All planes intersect at shell's center.
    In all panels $xyz$ are the Heliocentric Cartesian Galactic Coordinates.
    }
    \label{fig: 3D views dust shell}
\end{figure*}

\section{Results}
\label{sec: results}
In this section we explore the 3D structure of the Perseus-Taurus region, identifying 
the Per-Tau Shell, and the Tau Ring (\S \ref{sub: PerTau}-\ref{sub: shell profile}, and Table \ref{table}).
In \S \ref{sub: coompare to. 2D obs} we compare these structures with 2D observations, and explore observational tracers of past SN activity which may have led to the formation of the Per-Tau Shell.

\begin{table}[t!]
\caption{Summary of new structures identified}
\centering
\label{table}
 \begin{tabular}{ | l | l| l | } 
\hline
&   & \\
 Name & Center Coordinates & Radius \\
 \hline\hline 
   &   & \\
 Per-Tau Shell & $xyz = (-190, 65, -84)$ pc & $R_s=78$ pc \\
  & $lbd = (161^{\circ}.1, -22^{\circ}.7,  218 \ {\rm pc})$ &  \\
    &  ${\rm RA, Dec}= (53^{\circ}.0, \; 28^{\circ}.0)$ &  \\
    &   & \\
 Tau Ring & $xyz = (-174, 1, -44)$ pc & $a=39$ pc \\
   & $lbd = (179^{\circ}.5,  -14^{\circ}.2,  179 \ {\rm pc})$ & $b=26$ pc \\
   &${\rm RA, Dec} =(73^{\circ}.1 , \; 21^{\circ}.3)$&  \\
    &   & \\
 \hline 
  \end{tabular}
 \\{\vspace{0.2cm}}
$xyz$ are the Heliocentric Cartesian Galactic coordinates, and $lbd$ are Galactic lontitude, latitude and distance from the sun. 
$a$ and $b$ are the semi-major and semi-minor axis (see also Appendix \ref{app: ring}).
\end{table}

\subsection{Per-Tau Shell}
\label{sub: PerTau}

In Fig.~\ref{fig: 3D views dust shell} 
(see also the \href{https://faun.rc.fas.harvard.edu/czucker/Paper_Figures/sbialy/pertau_superbubble.html}{interactive figure}$^{(\ref{footnote: interactive figure})}$; for an augmented reality [AR] experience scan the QR code in Fig.~\ref{fig: 3D views dust shell} with a smartphone or a tablet)
we present iso-surfaces of gas at $n=5$ cm$^{-3}$ (grey) and 25 cm$^{-3}$ (color), for different viewing angles (see also Fig.~\ref{fig: 3D plus}).
The color shows the distance from the sun. 
The density iso-surfaces reveal that the diffuse gas in the region is organized in a near-spherical geometry.
The denser portion of the gas 
forms the Taurus and Perseus molecular clouds, lying on the near ($d \approx 150$ pc) and far ($d \approx 300$ pc) sides of the shell, respectively, where $d$ is distance from the Sun.

We model the Per-Tau Shell as a spherical shell with radius $R_s=78$ pc, as shown in Fig.~\ref{fig: 3D views dust shell} and \ref{fig: 3D plus} (see Table \ref{table} for the center coordinates).
The model's center was optimized by a visual inspection of the iso-surface data in 3D.
The model's radius was obtained by the position of the maximum in the radial density profiles, discussed in \S \ref{sub: shell profile} below.
In the bottom-right panel we show three density cuts through the shell's center, along the planes $xy$, $xz$, $yz$.
The density enhancements at $r=R_s$ are clearly evident.

We stress that the spherical-shell model only serves as an approximate representation of the real density structure.
In practice, the shell geometry is more complex. The shell is not a perfect symmetric sphere, and
it also exhibits significant density fluctuations (substructure).
This is manifested as the
``missing areas" in the shell's iso-surfaces (Fig.~\ref{fig: 3D views dust shell}, left), and the strong fluctuations in the density value in the density slices (Fig.~\ref{fig: 3D views dust shell}, right).
Such density fluctuations are expected to be present in the ISM due to supersonic turbulence and thermal instability 
\citep[e.g.,][]{Federrath2008, Kritsuk2017, Bialy2020a}.

\subsection{Tau Ring}
\label{sub: tau ring}
Another interesting feature revealed by the 3D dust observations is a prominent ring structure at the location of the Taurus cloud (see Fig.~\ref{fig: 3D views dust shell}), hereafter, the ``Tau Ring".
The Tau Ring is well represented by an ellipse with semi-major axis
$a=39$ pc
and semi-minor axis
$b=26$ pc, centered on
$\vec{r_0} \approx (-174,  1,  -44)$ pc, and viewed nearly edge-on along the LOS. 
In Appendix \ref{app: ring} we discuss Tau Ring's orientation and detailed structure.

\begin{figure}
    \centering
    \includegraphics[width=0.499\textwidth]{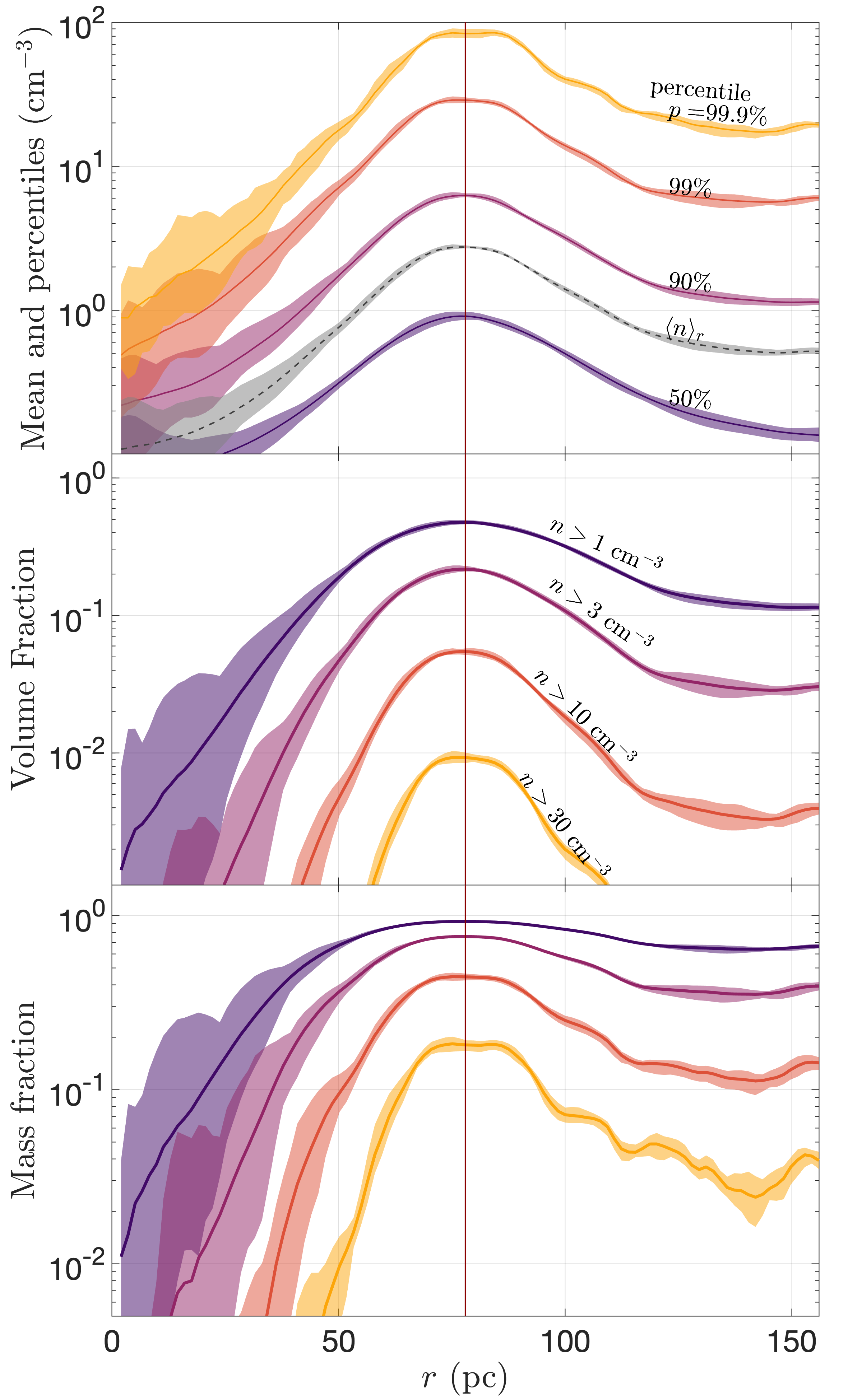}
    \caption{The Per-Tau Shell Radial Profile. The abscissa is the radial distance from the shell's center and the vertical line is its modeled radius (Table \ref{table}). {\bf Top}: The mean density (dashed), and different density percentiles (solid). {\bf Middle and Bottom}: Volume and mass filling fractions for gas with different densities.
    In all panels, all quantities are computed within thin concentric spherical shells (see Appendix \ref{sec: methodology}). 
    The shaded regions are the minimum-to-maximum variation across different dust-map samples, showing the uncertainty associated with the 3D data
    (see Appendix \ref{sec: methodology}).
    }
    \label{fig: shell profile}
\end{figure}

\subsection{Radial Density Profiles of the Per-Tau Shell}
\label{sub: shell profile}

In Fig.~\ref{fig: shell profile} we present the density profile of the Per-Tau shell, showing the radial mean density $\langle n \rangle_r$, density percentiles $n_p$, volume fractions $f_V$, and mass fractions $f_M$, as functions of the distance from the shell's center, $r$.
Each of these statistical quantities is computed within concentric spherical shells (Appendix \ref{sec: methodology}).

In Fig.~\ref{fig: shell profile} (top), $\langle n \rangle_r$ measures the mean radial trend in the region (averaging over many voxels at each radius).
To illustrate the density dispersion within each radius, the solid colored curves show different density percentiles.
The density percentiles span a significant range: for example, the 50-90 percentile
 spans 1 dex in density, and the 50-99.9 range
is as large as 2 dex in density.
The median density is below the mean, as expected for a lognormal density distribution.
Overall, the mean and the density percentiles show a clear trend:
increasing with $r$ at small radii; peaking at $R_s = 78$ pc, and then decreasing with $r$ at larger radii.

In the middle and lower panels, we show the volume and mass fractions, $f_V$ and $f_M$, of gas with various densities: $n > 1$, 3, 10, 30 cm$^{-3}$  (Appendix \ref{sec: methodology}).
Since $f_V$ is calculated within thin shells, it also approximately equals the area covering fraction of gas with density $>n$ at each radius.
All $f_V$ curves fall rapidly at small radii, within the shell void ($r \ll R_s$), as well as at higher radii ($r \gg R_s$), albeit more gradually.
Even at shell's peak, the denser gas occupies only a small fraction of the volume (and area).
However, in terms of mass fraction, the denser gas is significant.
For example, at the shell radius $n > 30$ cm$^{-3}$ gas accounts for
$\approx 20 \%$ of the mass, and 
 $\approx$ half the gas has density $> 10$ cm$^{-3}$.
This is in strong contrast with the central region (the void), which consists of mainly diffuse $n \ll 1$ cm$^{-3}$ gas.
The flat maximum (width $\approx 20$ pc) results from the Per-Tau Shell having a complex ovaloid shape rather than a perfect sphere (e.g., see Fig.~\ref{fig: 3D views dust shell}, lower-right panel).

In all panels, the shaded strips about each curve show the maximum and minimum values as obtained from each of the twelve samples presented in \citetalias{Leike2020a}, providing a measure of the uncertainty associated with \citetalias{Leike2020a}'s 3D dust map (Appendix \ref{sec: methodology}).
At small radii the spread is large as the statistics include only a small number of voxels, and thus the effect of fluctuations within the samples becomes significant. 
However, as $r$ increases these fluctuations become insignificant.

\begin{figure*}
    \centering
    \includegraphics[width=0.9\textwidth]{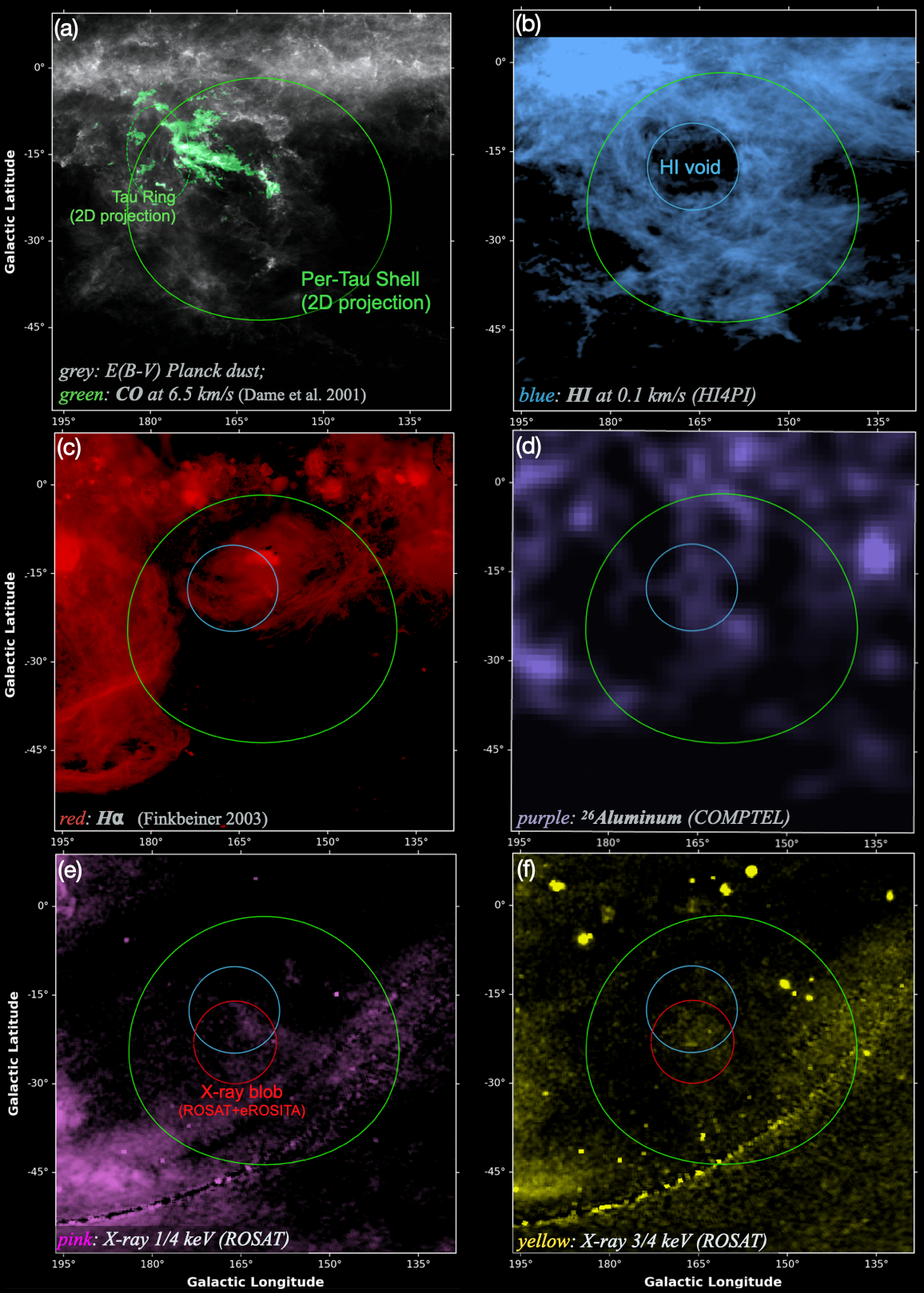}
    \caption{The Per-Tau region in different tracers:
    {\bf (a)}: Planck's $E(B-V)$ dust \citep{Abergel2014} and  \citet{Dame2001a}'s $^{12}$CO.
    2D projections of the Per-Tau Shell and the Tau Ring models are shown.
   {\bf (b)}: HI4PI's HI map \citep[][]{Bekhti2016}.
   {\bf (c)}: \citet{Finkbeiner2003}'s H$\alpha$ map.
   {\bf (d)}: COMPTEL's $^{26}$Al map  \citep[1.8 MeV $\gamma$-ray;][]{Diehl1995}.
      {\bf (e-f)}: ROSAT's X-ray 1/4. and 3/4 keV  \citep[][]{Snowden1997} (see also Fig.~\ref{fig: Xrays} for eROSITA's observations).
    The HI void and the X-ray emission blob (see also Appendix \ref{app: Xrays}) are indicated.
    The HI void, $^{26}$Al, H$\alpha$ and X-ray observations suggest recent SN activity in the region (see \S \ref{subsub: Al26, Xray and 21cm}).}
    \label{fig: 2D sky}
\end{figure*}

\subsection{2D observations}
\label{sub: coompare to. 2D obs}

\subsubsection{Dust and CO: Comparing 3D to 2D}
\label{subsub: Planck}
Fig.~\ref{fig: 2D sky}a shows the Planck $E(B-V)$ dust map tracing the integrated dust column density along the LOS, along with $^{12}$CO.
The 2D projections of the Per Tau Shell and the Tau Ring models are shown (Table \ref{table}). 
While the Tau Ring is partially traced by CO, the Per-Tau Shell is not identified in the 2D dust map or CO.
This is because these 2D maps probe the integrated column along the LOS,
and thus are much more sensitive to dense (high column density) structures.
The limb brightening that may have revealed the shell contour is not effective because
the large-scale structure of the shell consists of mainly diffuse gas ($n \sim 5$ cm$^{-3}$) and the associated column density is low, comparable to that of the ambient ISM.

Planck's $E(B-V)$ dust map is also shown in Fig.~\ref{fig:finding chart}.
Comparing the right and left panels we see that the 3D projected structures trace well
known features seen in Planck (see also Fig.~\ref{fig: 2d_3d} and \citealt{Zucker2021_topology}).
However, the 3D dust includes distance information, providing additional insights:
\begin{enumerate}
\item
{\it Taurus and Perseus:}
 The Taurus and Perseus molecular clouds are located at distances of $d\approx 150$ and 300 pc, from the sun (in agreement with \citealt{Zucker2020}), and are placed at the front and at the back sides of the Per-Tau Shell.
\item
{\it Tau Ring:} In a sky projection the Tau Ring is seen almost edge-on.
The near side of the Tau Ring connects with the main body of Taurus at $d \approx 150$ pc, whereas the farthest part extends to $d \approx 220$ pc.
\item
{\it The Fictitious Connection:} A filament seems to connect Taurus to Perseus.
This connection is only a coincidental projection effect, where in actuality the filament is located at the distance of Taurus, and does not physically connect to Perseus.
This coincidence is even more striking given that the connection not only occurs in 2D position, but also in velocity space (i.e., see $^{12}$CO, $v=6.5$ km/s channel map in Fig.~\ref{fig: 2D sky}). 
This emphasizes how misleading it is to infer 3D structures from PPV data alone.
\end{enumerate}

\subsubsection{Tracing SN feedback: HI, H$\alpha$, $^{26}$Al, and X-rays}
\label{subsub: Al26, Xray and 21cm}

Fig.~\ref{fig: 2D sky}b shows HI4PI's HI channel map ($v=-1.2$ km/s) of the Per-Tau region.
The prominent HI void (as indicated) was  studied previously by \citet{Sancisi1974}.
From the semi-circular arcs seen in the HI channel maps and the expansion signatures seen in $p-v$ diagrams, they concluded that the HI forms an expanding shell of radius $R\approx 20$ pc, potentially powered by previous SN which has exploded $\approx 1-4$ Myr ago
 \citep[see also][]{Lim2013, Shimajiri2019}.

Within the HI cavity, there is a peak in H$\alpha$ emission (Fig.~\ref{fig: 2D sky}c).  
The bright H$\alpha$ spot at $l\approx 162^{\circ}$, $b=-12^{\circ}$ is unrelated to the Per-Tau shell as it originates from the California nebula at a larger distance.
Ignoring the unfortunate California Nebula projection, Fig.~\ref{fig: 2D sky} shows that H$\alpha$ essentially fills the cavity evident in HI, consistent with the hypothesis of recent SN activity in the region.

Fig.~\ref{fig: 2D sky}d shows an $^{26}$Al map.
$^{26}$Al is injected into the ISM mainly by massive stars via core collapse SN and in Wolf-Rayet (WR) winds \citep{Prantzos1996, Diehl2004}.
We identify a blob of enhanced $^{26}$Al emission near the center of the Per-Tau Shell ($l \approx 168^{\circ}$, $b\approx -25^{\circ}$).
Assuming the blob is located at the distance of the shell's center ($d=225$ pc), and using a distance-flux scaling based on observations of the Orion-Eridanus region \citep{Diehl2002}, we obtain an $^{26}{\rm Al}$ mass of $\approx 6\times 10^{-5}$ $M_{\odot}$, consistent with production by WR stars or SN \citep{Limongi2006}. 
As there are no living WR stars within the shell, and given that $^{26}$Al's half-lifetime is 0.7 Myrs, the observed emission suggests recent SN activity over the last few Myrs.

Fig.~\ref{fig: 2D sky}(e-f) show ROSAT's 1/4 and 3/4 keV diffuse X-ray emission.
We find an enhancement of X-ray emission near Per-Tau shell's projected center (red circle).
We have also identified an X-ray enhancement at a similar position in eROSITA's all-sky map (see Appendix \ref{app: Xrays}).
This diffuse X-ray emission (radius $\approx 7^{\circ}$) points towards the existence of hot gas potentially produced by past SN(e) on a timescale of a Myr \citep[or less;][]{Krause2014}.
Given the proximity to the HI void, the X-ray blob and the HI void may have resulted from the same SN event.

\section{Discussion}
\label{sec: discussion}

We have shown that the ISM in the Perseus-Taurus region has the structure of a large nearly spherical shell of radius $R_s \approx 78$ pc, encompassing the molecular clouds Taurus and Perseus, which are located at the near and far sides of the shell, at $d \approx 150$ pc, and $\approx 300$ pc.
The Taurus cloud also extends to larger distances, forming an extended ellipsoidal ring with semi-major and semi-minor axis $a\approx 39$ pc, $b\approx 26$ pc (Table \ref{table} and Appendix \ref{app: ring}).

Recently, \citet{Doi2021} also mentioned the presence of a large dust cavity, as seen in the previous, lower resolution \citet{Leike2019} map.
However, as their focus is on the magnetic fields in the region, they did not study the 3D density structure in detail as is done here, 
nor did they explore potential mechanisms for forming this structure.
As they show, the bimodal distribution of polarization angles supports the idea that Perseus and Taurus are well separated along the LOS.

In the remainder of this section we discuss a physical scenario for the formation of the Per-Tau Shell via gas compression and cooling in the swept-up shell of an expanding SB, and the implications for star formation (positive) feedback. 
We also estimate the shell's age, and momentum and energy sources.

\subsection{A Formation Scenario for the Per-Tau Shell}
\label{sub: formation scenario}
The shell's near-spherical geometry, prominent dust free cavity, and large extent suggest a scenario in which the Per-Tau Shell has formed via multiple SN episodes, driving an expanding SB, sweeping up ISM into an extended shell.
The HI, $^{26}$Al and X-ray observations discussed in \S \ref{subsub: Al26, Xray and 21cm} further support this scenario.
As discussed in \S \ref{sec: intro}, expanding shells may promote the formation of cold and molecular gas, and star formation.
Indeed, the Per-Tau shell includes the
Perseus and Taurus clouds, which are dense, molecule-rich gas clouds, which are actively forming stars.

However, there are no signatures of shell expansion:
the Perseus and Taurus clouds are located on two opposite sides of the shell (along the LOS), but their CO radial velocities are similar to within a few km/s.
The lack of expansion is also suggested by the 3D space motions of young stars in the two clouds \citep[][]{Ortiz-Leon2018, Galli2019}. Thus, in this scenario, the Per-Tau Shell must be an old remnant, which has already decelerated below the turbulent velocity dispersion of the ISM, $v_{\rm turb} \approx 7$ km/s, and is now dominated by gas dynamics of the surrounding ISM.
Indeed, the ISM is expected to be abundant in such old bubbles due to the large volume they occupy and the long time spent in the evolved stage \citep[][]{McKee1977, Cox2005}.
The evolved stage is also the time at which there is higher probability for observing dense molecular gas, and star-formation in the shell, as
sufficient time has passed to allow cooling to low temperatures, molecule-formation, and gravitational collapse.

\subsection{Timescales}
\label{sub:  age estimation}

We can obtain an estimate for the Per-Tau Shell's age from its observed radius and lack of radial expansion.
Theoretical models of an expanding SN remnant into a uniform medium find $R \propto t^{\eta}$ with
$\eta = 0.25-0.3$ at late stages \citep[at the snowplow phase;][]{Cioffi1988, Kim2015, Haid2016a}.
For a more realistic scenario of a SB powered by multiple SN $\eta \approx 0.5$ \citep[][see, e.g., Fig.~6 in the latter paper]{kim2017b,El-Badry2019}.
The expansion velocity evolves as $v_{\rm exp}=\eta (R/t)$,
and the shell age  is
\begin{equation}
\label{eq: t_min}
    t = \eta \frac{R}{v_{\rm exp}} = 5.5 \left(\frac{\eta}{0.5}\right)\left(\frac{R}{78 \ {\rm pc}} \right)\left(\frac{v_{\rm exp}}{7 \ {\rm km/s}} \right)^{-1} \ {\rm Myr} \ .
\end{equation}
Here we normalized $R$ to the observed shell's radius, and the expansion velocity to $v_{\rm turb}=7$ km/s,
a typical value for the turbulent velocity dispersion (also comparable to the ambient WNM sound speed).
Setting $v_{\rm exp}=v_{\rm turb}$ in Eq.~(\ref{eq: t_min}) we get a {\it lower limit} for the shell's age, $t_{\rm min} \approx 6$ Myr.
This is a lower limit because, in practice the shell does not expand today, and the point at which it decelerated to $v=v_{\rm turb}$ has occurred in the past, an unknown amount of time ago.

On the other hand, the shell cannot be too old, as otherwise the surrounding ISM would have filled the void inside the shell.
The characteristic timescale for this process is the turbulent turnover time,
\begin{equation}
\label{eq: t_max}
    t = \phi \frac{R}{v_{\rm turb}} = 21.8 \left(\frac{\phi}{2}\right) \left(\frac{R}{78 \ {\rm pc}} \right)\left(\frac{v_{\rm turb}}{7 \ {\rm km/s}} \right)^{-1} \ {\rm Myr} \ ,
\end{equation}
where $\phi$ is a factor of order unity that accounts for the uncertainty in this simplistic description.
For a conservative upper limit we adopt $\phi=2$, giving $t_{\rm max}\approx 22$ Myrs.
The Per-Tau Shell age is thus bracketed $t_{\rm age} \approx (6-22)$ Myrs.

Our derived age is long compared to the stellar ages of Perseus and Taurus \citep{Bally2008, Luhman2018}, consistent with the scenario where the clouds and subsequent stars have formed in the shell.
$t_{\rm age}$ is also long compared to the timescales of feedback traced by the
 HI, $^{26}$Al and X-rays observations (see \S \ref{subsub: Al26, Xray and 21cm}). 
This suggests that not one, but several
SNe have probably contributed to the inflation of the Per-Tau Shell.
In this picture, $t_{\rm age}$ is the time when the first SN(e) went off, whereas HI, $^{26}$Al and X-rays trace more recent feedback activity, within the last $\sim 1$ Myr, in line with 
SN clustering \citep{Wit2004, Fielding2018}.

\subsection{Momentum and Energy Sources}
\label{sub:  age estimation}
Simulations of clustered SNe exploding find that the momentum imparted into the ISM, per SN, is  $\hat{p} = (3-6) \times 10^5$ M$_{\odot}$ km/s for SN frequency of 1/Myr \citep{El-Badry2019}.
For the Per-Tau Shell, we calculate the swept-up mass, $M$, by integrating the density (as given by the 3D density cube; \S \ref{sec: data}) inside a sphere of radius $R$, and multiplying it by a mean particle mass $\langle m \rangle=1.4 m_{\rm H}$.
To have a handle on the uncertainty we perform the calculation for two different radii: $R=94$ pc and 117 pc, corresponding to 1.2 and 1.5 $R_s$ where $R_s=78$ pc is the density-peak radius.
These values are motivated by the gradual decrease of the radial profiles in Fig.~\ref{fig: shell profile}. 
We obtain $M = (1.7-3.2) \times 10^5$ M$_{\odot}$, respectively.
The associated momentum is
\begin{equation}
\label{eq: p_tot}
    p = M v_{\rm exp} = (1.2-2.3) \times 10^6 \left( \frac{v_{\rm exp}}{7 \ {\rm km/s}} \right) \ {\rm M_{\odot} \ km/s} \ .
\end{equation}
Dividing by $\hat{p}$, we get that the corresponding number of SNe is
\begin{equation}
\label{eq: N_SN}
    N_{\rm SN} = \frac{p}{\hat{p}} = (2-8) \left( \frac{v_{\rm exp}}{7 \ {\rm km/s}} \right)  \ .
\end{equation}
The (2-8) range corresponds to the maximum mutual uncertainty range for $M(r)$ and $\hat{p}$.
In practice, since the expansion velocity has already decreased below the $7$ km/s level, Eqs.~(\ref{eq: p_tot}-\ref{eq: N_SN}) are upper limits on $p$ and $N_{\rm SN}$.

The total mass of the star cluster required to produce these SNe can be estimated by assuming an IMF and a distribution of massive star lifetimes. 
For the \citet{Chabrier2003} IMF, a star cluster of mass $M_\star$ produces in total $\sim0.01(M_\star/M_\odot)$ SNe. Assuming the stellar lifetime-mass relation for solar metallicity from \citet{Raiteri1996}, we estimate that $\sim 15\%-65\%$ of these SNe explode by $6-22$ Myr, the age of the shell. For $N_{\rm SN} \sim 5$ (the average from Eq.~\ref{eq: N_SN}), these numbers correspond to a total star cluster mass of $M_\star \sim 800 - 3300\;M_\odot$. 

Note however, that these estimations for $N_{\rm SN}$ and $M_{\star}$ may be considered upper limits, since they are based on simulations where $\hat{p}$ is likely underestimated. 
First, these simulations lack stellar winds and radiation pressure which can inject momentum into the ISM at a rate comparable with SNe \citep[e.g.,][]{Agertz2013}. 
Second, they do not include the nonthermal pressure exerted by cosmic rays, which can boost the momentum injected per SN by a factor of a few \citep[][see their Fig.~3]{Diesing2018}. 
Finally, the momentum per-SN may be underestimated due to excessive numerical 
mixing of the superbubble material with the surrounding ISM \citep[e.g.,][]{Gentry2019}.

Where are the remnants of the stellar cluster where the powering SN went off?
Two recent studies have identified young populations inside the Per-Tau shell, with \textit{Gaia}.
\cite{Pavlidou2021} reported the discovery of five new stellar groups in the vicinity of the Perseus cloud ages between 1 - 5 Myr.
\cite{Kerr2021} characterized all young stars in the local neighborhood. 
Their results suggest that there are at least two young  (age $<$ 20 Myr) populations located inside the Per-Tau shell. 
 Pre-\textit{Gaia}, \cite{Mooley2013-lw} identified several B- and A-stars towards the Taurus clouds, consistent with the existence of a young population inside the Per-Tau, likely the more massive counterparts of the recently discovered Gaia populations. 
 A dedicated study of young stars inside Per-Tau is currently missing, and the insights brought by the finding in this paper make such a study a priority for the near future.

Given the large uncertainty in the estimate of $N_{\rm SN}$, an alternative scenario that does not require a star cluster within Per-Tau is that the Per-Tau Shell has been produced by only a single SN explosion.
Such a SN can result from an O or B star dynamically ejected from a nearby star-forming region.
A more exotic possibility is that the Per-Tau bubble is a relic of a previous ultraluminous X-ray source (ULX).
ULXs are often surrounded by hot ionized gas bubbles, thought to be powered by strong winds and radiation pressure
\citep[e.g.,][]{Pakull2001,Kaaret2017,Sathyaprakash2019, Fabrika2021}. 
However, since ULXs are extraordinarily rare, this possibility is rather unlikely.

\section{Conclusions}
\label{sec: conclusions}
In this paper we explored the 3D structure of the Perseus and Taurus region.
Our conclusions are
\begin{enumerate}
    \item  The Perseus and Taurus molecular clouds are not independent, but are part of a larger structure, the ``Per-Tau Shell", an extended shell of radius 78 pc, centered in-between the two clouds at a distance $d= 218$ pc.
    \item The 3D observations further reveal a prominent ring structure connecting to Taurus, the ``Tau Ring", with semi-major and semi-minor axes, $a=39$ pc, $b=26$ pc, oriented nearly edge-on and is partially traced by CO.
    \item The densest structures in the 3D map have good correspondence with known features in the 2D dust and CO, but diffuse structures (such as the Per-Tau Shell) are not identified in 2D as the integration along the LOS results in significant blending with the ambient ISM.
    \item H$\alpha$, Soft X-ray, and $^{26}$Al enhancements, as well as a void of HI, are seen in different locations within the Per-Tau Shell, suggesting recent SN activity.
\end{enumerate}
Based on these observations we propose a formation scenario for the Per-Tau Shell by  multiple SNe and other forms of stellar feedback activity occurring within the last $\approx 6-22$ Myrs. This feedback inflated an expanding shell of compressed cooled gas, potentially leading to the formation of the Taurus and Perseus molecular clouds.
This points to the positive aspect of stellar/SN feedback, promoting cloud formation and star formation in expanding superbubbles.

\appendix

\section{A. Methods}
\label{sec: methodology}

{\bf Radial Mean Density:}
To characterize the structure of the Per-Tau shell we calculate the radial-mean density profile, defined as:
\begin{equation}
\label{eq: n(r)}
        \langle n \rangle_r(r) \equiv \frac{\sum_{r' \in (r-\delta r/2, r+\delta r/2)} \; n(r')}{\sum_{r' \in (r-\delta r/2, r+\delta r/2)}1}\ .
\end{equation}
Here $r$ is the radial distance from Per-Tau shell's center.
At each radius, the density is averaged over a thin radial shell of radius $r$ and width $\delta r$.
We choose $\delta r=10$ pc, large compared to the data resolution (1 pc).
This (1) reduces inaccuracies resulting from the misalignment of the Cartesian grid of the data with the spherical geometry of the averaging window, and (2) allows having a sufficiently large number of voxels for calculating statistics. 
We also experimented with other values for $\delta r=5-20$, and verified that our results are insensitive to the exact value of $\delta r$.
We compute $\langle n \rangle_r(r)$ as a function of $r$, for $r$ ranging from 0 (shell's center) to 140 pc.

{\bf Radial Percentiles and Volume and Mass Fractions:}
While  $\langle n \rangle_r(r)$ is useful for exploring the general trend with radius, it gives only the mean density 
as a function of radius and does not include information on the density dispersion within each radius.
To characterize the shell profile in more detail, we calculate density percentiles, and the volume and mass fractions.
The density percentile $n_{\rm p}(r)$ is computed as the density below which gas occupies the volume fraction $p$ in a radial shell with radius $r$. For example, $n_{p=0.95}(r)=10$ cm$^{-3}$ means that at radius $r$, gas with $n < 10$ cm$^{-3}$ fills 95\% of the volume, whereas dense $n \geq 10$ cm$^{-3}$ gas is rare and occupies only the remaining 5\% of volume.
In a similar manner, at any given radius $r$, the volume and mass fractions, $f_V(r; n')$ and $f_M(r; n')$, respectively, are the fractions of the volume and mass at that radius, occupied with gas denser than $n'$.
Similarly to the calculation of $\langle n \rangle_r(r)$,
we calculate
$n_{\rm p}$, $f_V$ and $f_M$, within thin radial shells of radii $r$ (ranging from 0 to 140 pc), and width $\delta r=10$ pc.

{\bf Uncertainties in the 3D Dust Map:}
\citetalias{Leike2020a} presented twelve posterior samples for the dust opacity density, representing possible realizations of the underlying 3D dust distribution.
While the samples are statistically similar, they differ in the density they predict on a voxel-by-voxel basis.
For the radial profiles discussed above.
we repeat our calculation for each of the twelve samples provided by \citetalias{Leike2020a}.
In \S \ref{sec: results}, we present the mean profiles, averaged over the sample results, as well as the minimum-maximum ranges for the twelve samples.
The dispersion over the samples provides a handle on the uncertainty associated with the reconstruction method adopted by \citetalias{Leike2020a}.
Otherwise, where statistical calculations are not involved, we
utilize the mean dust map in which the density is averaged over the samples.

\begin{figure*}[t]
    \centering
    \includegraphics[width=1\textwidth]{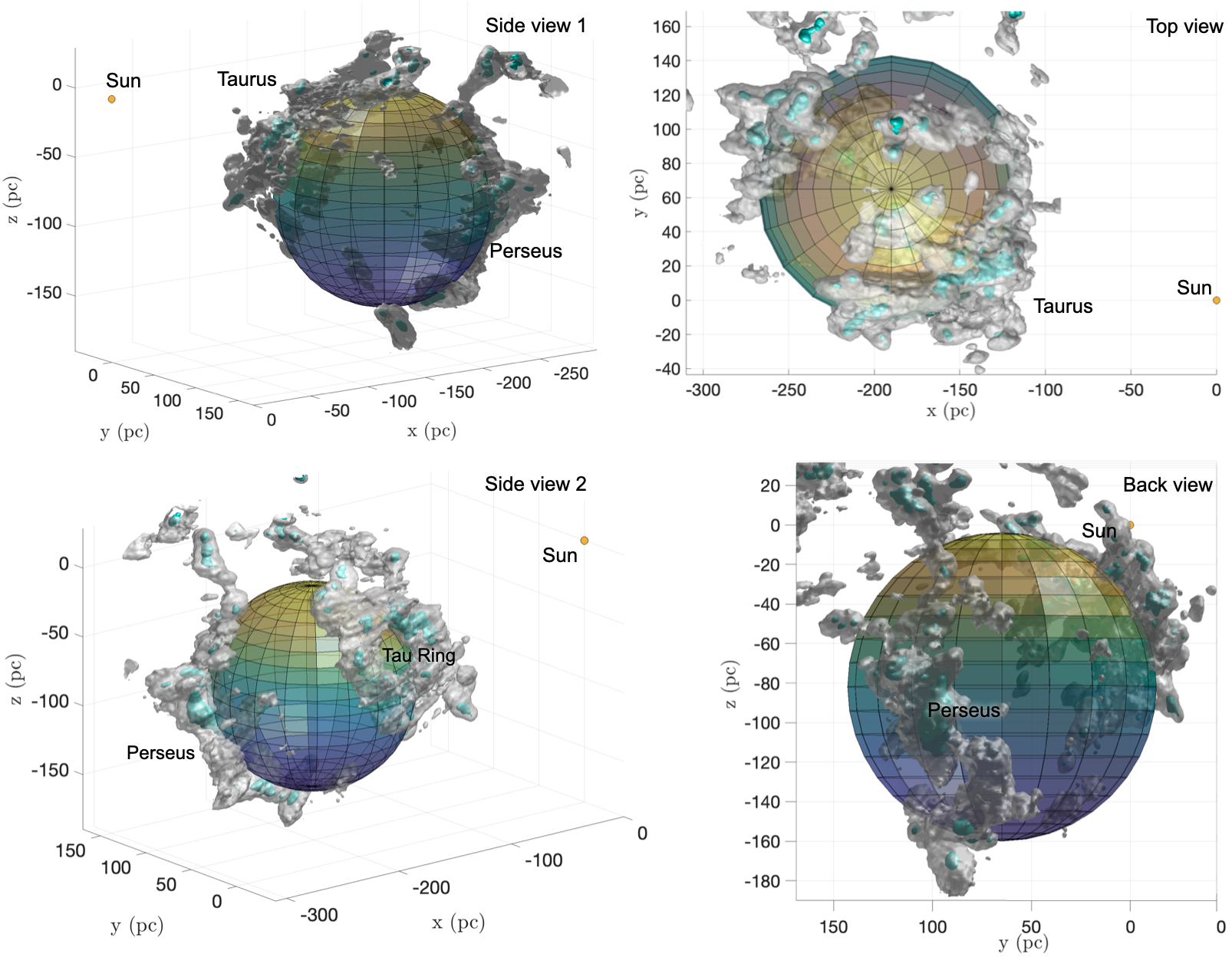}
            \includegraphics[width=1\textwidth]{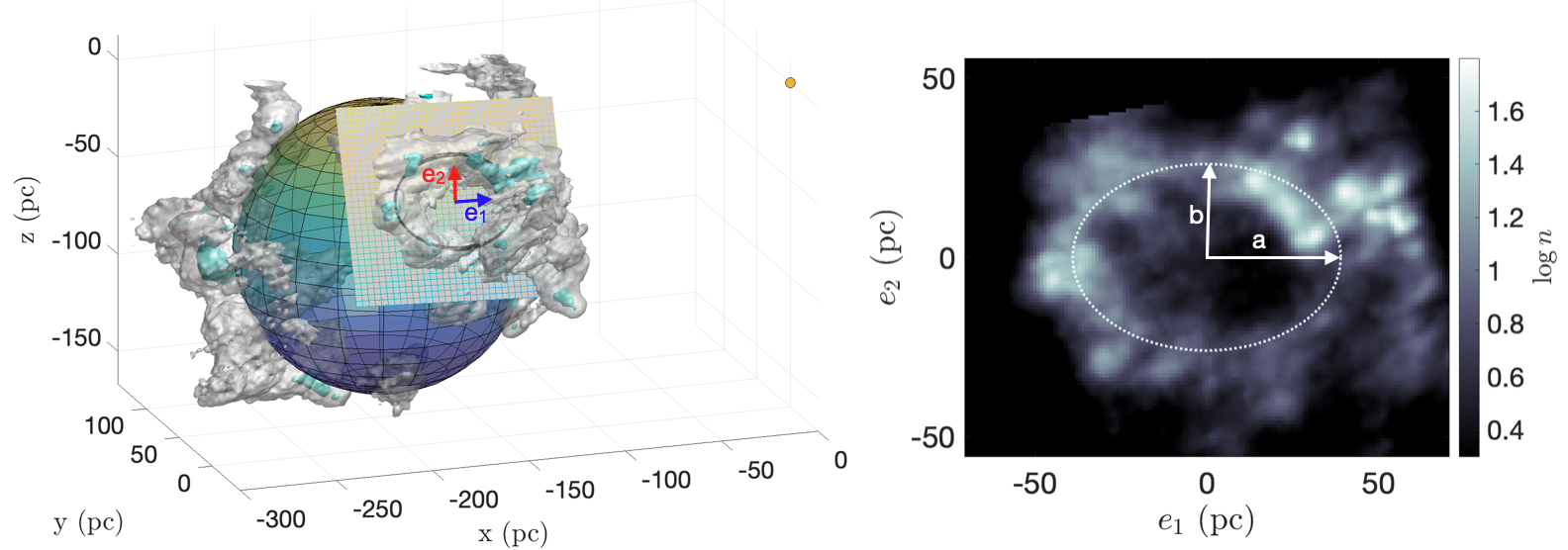}
    \caption{3D view of the Perseus Taurus region in 3D dust. 
    {\bf Top Panels}: Shown are iso-surfaces at $n=5$ cm$^{-3}$ (grey) and $n=25$ cm$^{-3}$ (cyan) levels. The colored sphere is the sphere model of the Per-Tau shell (Table \ref{table}). 
    {\bf Bottom Panels:}
    The Tau Ring. A plane cutting through the Tau Ring (left), the basis vectors spanning the plane (\ref{eq: basis vectors}) are also shown. On the Right we show a density slice through the $\hat{e}_1-\hat{e}_2$ plane. The modeled ellipse with semi-major axis $a=39$ pc and and semi-minor axis $b=26$ pc is shown.
    }
    \label{fig: 3D plus}
\end{figure*}

\begin{figure*}[t]
    \centering
    \includegraphics[width=1\textwidth]{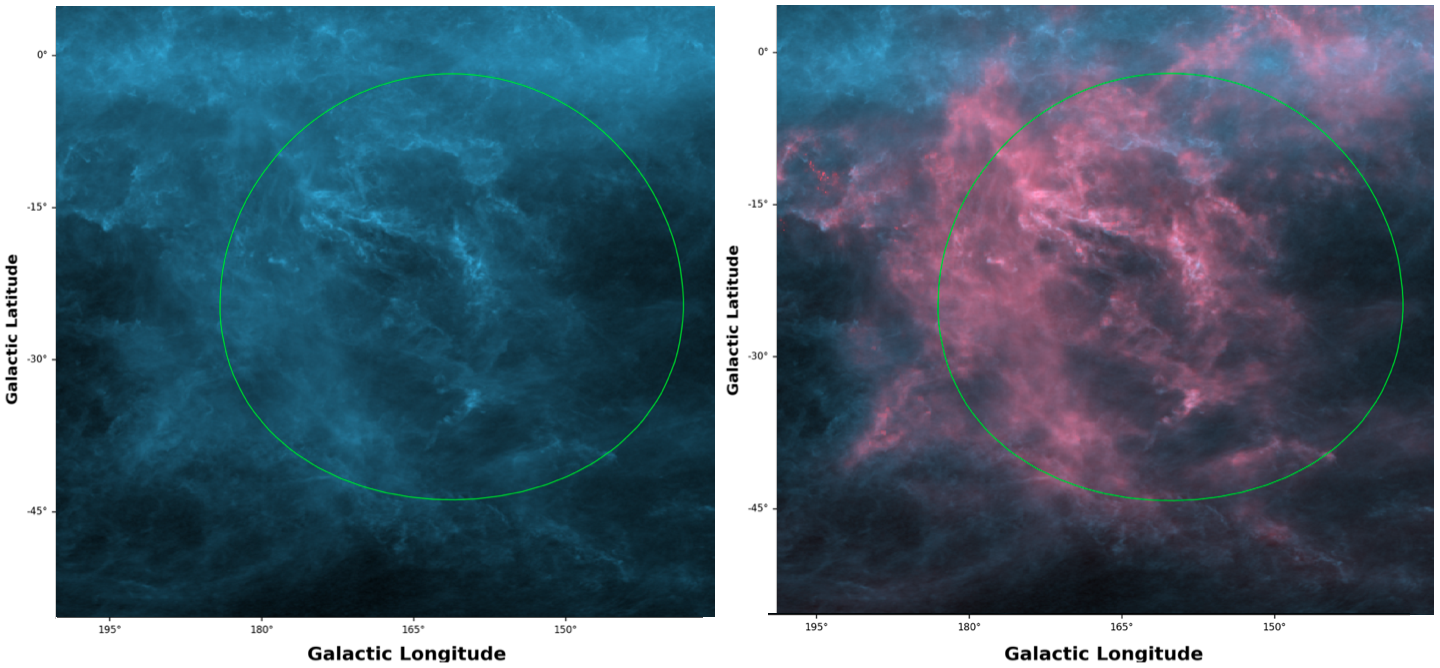}
    \caption{Comparing 3D dust to 2D dust. Left: The Planck $E(B-V)$ dust map (in blue) tracing the total dust column density integrated along the LOS. Right: The Planck $E(B-V)$ dust map overlaid with the projection of the 3D dust map (in red) on the plane of the sky. 
    The two maps are in excellent agreement. 
    The 3D dust includes all dust within the \citetalias{Leike2020a} 3D map limits (e.g., the California Nebula which is located at a larger distance is excluded and does not show in red).
    }
    \label{fig: 2d_3d}
\end{figure*}

\begin{figure*}[t]
    \centering
\includegraphics[width=1\textwidth]{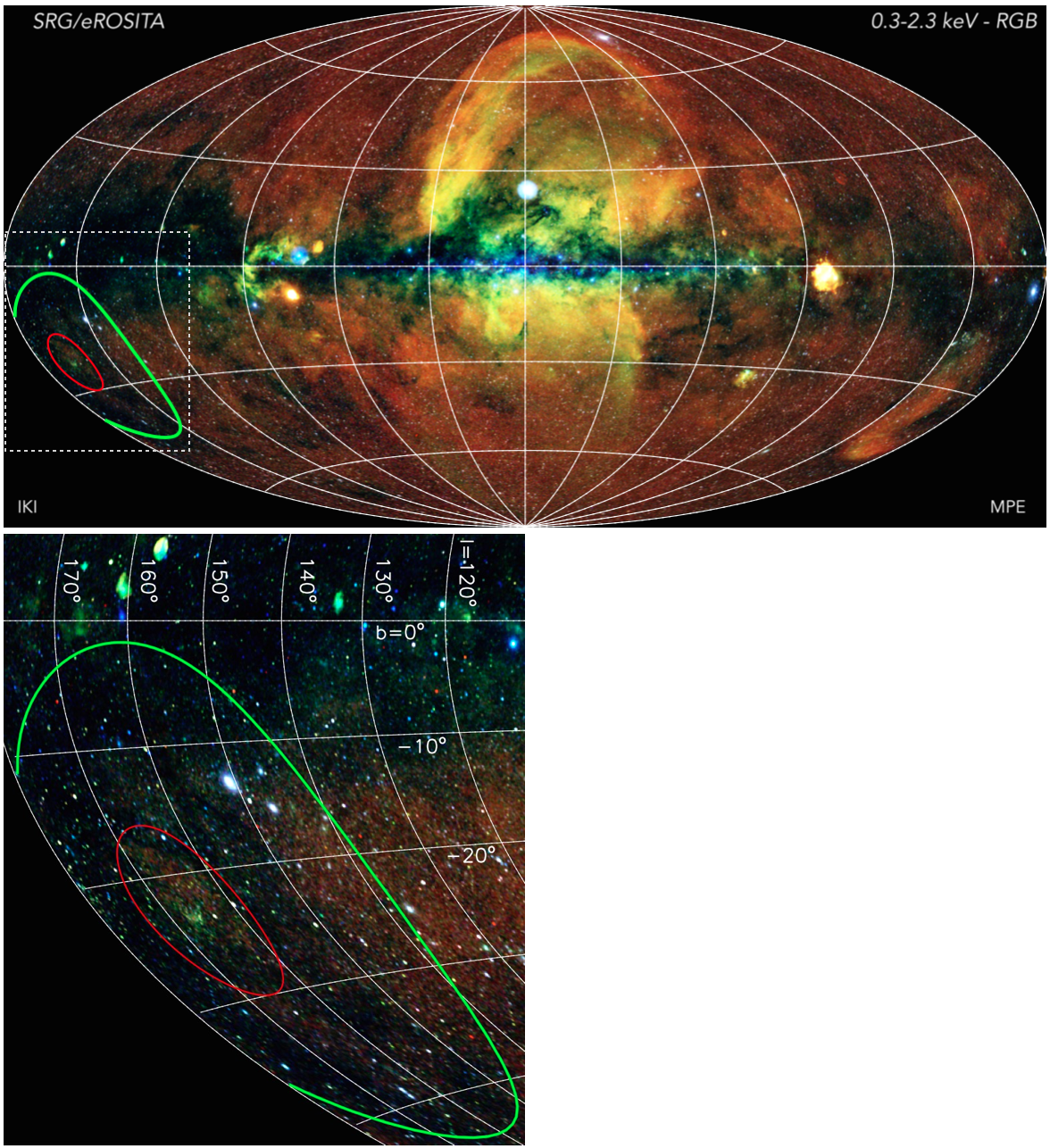}
    \caption{eROSITA's soft X-ray image of the Per-Tau region (originally published online$^{\ref{footnote: Xray}}$).  
    The RGB colors correspond to  X-ray energies, from 0.2 keV (red) to 2.3 keV (blue).
    The green oval is the projected Per-Tau shell outline (the distorted shape is a result of the Aitoff map projection of eROSITA's all sky map).  
    The red oval within the Per-Tau shell, highlights a region of enhanced X-ray emission, which is also seen in the ROSAT X-ray data (see Fig.~\ref{fig: 2D sky} panels e-f). 
        }
        \label{fig: Xrays}
\end{figure*}

\section{B. The Tau Ring and Additional 3D Visualizations}
\label{app: ring}
In Fig.~\ref{fig: 3D plus} we show additional 3D visualizations of the dust in the Perseus-Taurus region as seen from different observational orientations. Our model of the Per-Tau shell is also shown (Table \ref{table}).
In the bottom-left panel we show a 2D plane that cuts through the Tau Ring.
The plane is spanned by the two basis vectors:
\begin{equation}
\label{eq: basis vectors}
    \hat{e}_1 = (0.923, \, 0.353,  \, -0.154) \; , \; \hat{e}_2 = (0.005,  \,  0.302, \,   0.841) \ ,
\end{equation}
in Heliocentric Cartesian Galactic coordinates. The origin of the basis vectors, and the Tau ring, is located at
\begin{equation}
\label{eq: ring center}
    \vec{r_0} = (-174, \, 1, \, -44) \ {\rm pc} \ .
\end{equation}
The $\hat{e}_1-\hat{e}_2$ basis vectors are shown as the blue and red vectors in the lower-left panel of Fig.~\ref{fig: 3D plus}.
Any point on the $\hat{e}_1-\hat{e}_2$ plane is described by $\vec{r} = \hat{e}_1 s+\hat{e}_2 t+\vec{r_0}$ where $s,t$ are some real numbers.
In the bottom-right panel we show a density cut through the $\hat{e}_1-\hat{e}_2$ plane. 
We model the Tau Ring as an ellipse with semi-major axis $a=39$ pc and and semi-minor axis $b=26$ pc.
The basis vectors $\hat{e}_1-\hat{e}_2$ were constructed to be parallel with the ellipse's semi-major and semi-minor axis, respectively.
 Thus, Tau Ring's 3D orientation is fully defined by $\hat{e}_1, \hat{e}_2$ (Eq.~\ref{eq: basis vectors}).
 
\section{C. Further 2D-3D dust Comparisons}
\label{app: 2d-3d comparison}
In Fig.~\ref{fig: 2d_3d} we show a comparison of 2D and 3D dust. In the left panel we show the Planck $E(B-V)$ dust map in blue. In the right panel we overlay Planck's map with a projected 3D dust map (in red).
The two are in excellent agreement.
The advantage of 3D dust is that it includes information on the distance, so we can exclude foreground and background material in the 3D projected maps. For example, the map shown in Fig.~\ref{fig: 2d_3d} does not show the background California Nebula.

\section{D. e-ROSITA X-ray Observations}
\label{app: Xrays}
In Fig.~\ref{fig: Xrays} we show eROSITA's soft X-rays all-sky map (originally published online\footnote{\label{footnote: Xray}\href{https://www.mpe.mpg.de/7461761/news20200619}{https://www.mpe.mpg.de/7461761/news20200619}}), and a zoom-in onto the region of the Per-Tau shell.
The green oval is the projection of the Per-Tau shell (the distorted shape is due to map projection effects).
We identify a blob of enhanced X-ray emission within the Per-Tau Shell in both ROSAT and eROSITA. 
This is shown as the red oval in Fig.~\ref{fig: Xrays}, as well as in panels (e,f) in Fig~\ref{fig: 2D sky}.


\acknowledgements
The authors thank Roland Diehl for making the COMPTEL $\gamma$-ray map accessible to us, Peter Williams for helping with coordinate projections of the ROSAT data, and the anonymous referee for useful comments. 
SB.~thanks Drummond Fielding, Shu-ichiro Inutsuka, Munan Gong, and Enrique Vazquez-Semadeni for insightful discussions.
The authors are thankful to Eugene Belyaev and the \href{https://www.delightex.com/}{Delightex} team for developing the augmented reality (AR) figure. The authors also thank Ian Masson for developing functionality to enable export of the data into AR-compatible formats. 
SB acknowledges the Institute for Theory and Computations (ITC) at the Harvard-Smithsonian Center for Astrophysics for financial support.
The visualization, exploration, and interpretation of data presented in this work was made possible using the \texttt{glue} visualization software, supported under NSF grant numbers OAC-1739657 and CDS\&E:AAG-1908419. 
The interactive component of Figure 2 was created using the visualization software (\url{plot.ly}).


\begin{thebibliography}{79}
\expandafter\ifx\csname natexlab\endcsname\relax\def\natexlab#1{#1}\fi

\bibitem[{Abergel {et~al.}(2014)Abergel, Ade, Aghanim, Alves, Aniano,
  Armitage-Caplan, Arnaud, Ashdown, \& et~al. .}]{Abergel2014}
Abergel, A., Ade, P.~A., Aghanim, N., Alves, M.~I., Aniano, G.,
  Armitage-Caplan, C., Arnaud, M., Ashdown, M., \& et~al. . 2014, A{\&}A, 571,
  1

\bibitem[{Agertz {et~al.}(2013)Agertz, Kravtsov, Leitner, \&
  Gnedin}]{Agertz2013}
Agertz, O., Kravtsov, A.~V., Leitner, S.~N., \& Gnedin, N.~Y. 2013, ApJ, 770

\bibitem[{Audit \& Hennebelle(2005)}]{Audit2005}
Audit, E. \& Hennebelle, P. 2005, A{\&}A, 433, 1

\bibitem[{Bally {et~al.}(2008)Bally, Walawender, Johnstone, Kirk, \&
  Goodman}]{Bally2008}
Bally, J., Walawender, J., Johnstone, D., Kirk, H., \& Goodman, A. 2008, Handb.
  Star Form. Reg., 4

\bibitem[{Bekhti {et~al.}(2016)Bekhti, Fl{\"{o}}er, Keller, Kerp, Lenz, Winkel,
  Bailin, Calabretta, Dedes, Ford, Gibson, Haud, Janowiecki, {W Kalberla},
  Lockman, McClure-Griffiths, Murphy, Nakanishi, Pisano, \&
  Staveley-Smith}]{Bekhti2016}
Bekhti, B., Fl{\"{o}}er, L., Keller, R., Kerp, J., Lenz, D., Winkel, B.,
  Bailin, J., Calabretta, M.~R., Dedes, L., Ford, H.~A., Gibson, B.~K., Haud,
  U., Janowiecki, S., {W Kalberla}, P.~M., Lockman, F.~J., McClure-Griffiths,
  N.~M., Murphy, T., Nakanishi, H., Pisano, D.~J., \& Staveley-Smith, L. 2016,
  A{\&}A, 594, 116

\bibitem[{Bialy \& Burkhart(2020)}]{Bialy2020a}
Bialy, S. \& Burkhart, B. 2020, ApJ, 894, L2

\bibitem[{Bialy \& Sternberg(2019)}]{Bialy2019}
Bialy, S. \& Sternberg, A. 2019, ApJ, 881, 160

\bibitem[{Brown {et~al.}(2018)Brown, Vallenari, Prusti, J, \& et~al.
  {Babusiaux, C}}]{Brown2018}
Brown, A. G.~A., Vallenari, A., Prusti, T., J, D. B. J.~H., \& et~al.
  {Babusiaux, C}. 2018, A{\&}A, 616, A1

\bibitem[{Chabrier(2003)}]{Chabrier2003}
Chabrier, G. 2003, Publ. Astron. Soc. Pacific, 115, 763

\bibitem[{Chen {et~al.}(2019)Chen, Huang, Yuan, Wang, Fan, Xiang, Zhang, Tian,
  \& Liu}]{Chen2019}
Chen, B.~Q., Huang, Y., Yuan, H.~B., Wang, C., Fan, D.~W., Xiang, M.~S., Zhang,
  H.~W., Tian, Z.~J., \& Liu, X.~W. 2019, MNRAS, 483, 4277

\bibitem[{Cioffi {et~al.}(1988)Cioffi, Mckee, \& Bertschinger}]{Cioffi1988}
Cioffi, D.~F., Mckee, C.~F., \& Bertschinger, E. 1988, ApJ, 334, 252

\bibitem[{Cox(2005)}]{Cox2005}
Cox, D.~P. 2005, ARA{\&}A, 43, 337

\bibitem[{Dame {et~al.}(2001)Dame, Hartmann, \& Thaddeus}]{Dame2001a}
Dame, T.~M., Hartmann, D., \& Thaddeus, P. 2001, ApJ, 547, 792

\bibitem[{Dawson(2013)}]{Dawson2013}
Dawson, J.~R. 2013, Publ. Astron. Soc. Aust., 30, e025

\bibitem[{Dawson {et~al.}(2011)Dawson, Kawamura, Mizuno, Onishi, Mizuno, \&
  Fukui}]{Dawson2011}
Dawson, J.~R., Kawamura, A., Mizuno, N., Onishi, T., Mizuno, A., \& Fukui, Y.
  2011, ApJ, 728, 127

\bibitem[{Dawson {et~al.}(2015)Dawson, Ntormousi, Fukui, Hayakawa, \&
  Fierlinger}]{Dawson2015}
Dawson, J.~R., Ntormousi, E., Fukui, Y., Hayakawa, T., \& Fierlinger, K. 2015,
  ApJ, 799, 64

\bibitem[{Diehl(2002)}]{Diehl2002}
Diehl, R. 2002, New Astron. Rev., 46, 547

\bibitem[{Diehl {et~al.}(2004)Diehl, Cervi{\~{n}}o, Hartmann, \&
  Kretschmer}]{Diehl2004}
Diehl, R., Cervi{\~{n}}o, M., Hartmann, D.~H., \& Kretschmer, K. 2004, New
  Astron. Rev., 48, 81

\bibitem[{Diehl {et~al.}(1995)Diehl, Dupraz, Bennett, \& Bloemen}]{Diehl1995}
Diehl, R., Dupraz, C., Bennett, K., \& Bloemen, H. 1995, A{\&}A, 298, 445

\bibitem[{Diesing \& Caprioli(2018)}]{Diesing2018}
Diesing, R. \& Caprioli, D. 2018, Phys. Rev. Lett., 121, 91101

\bibitem[{Doi {et~al.}(2021)Doi, Hasegawa, Bastien, Tahani, Arzoumanian,
  Coud{\'{e}}, Matsumura, Sadavoy, Hull, Shimajiri, Furuya, Johnstone, Plume,
  Inutsuka, Kwon, \& Tamura}]{Doi2021}
Doi, Y., Hasegawa, T., Bastien, P., Tahani, M., Arzoumanian, D., Coud{\'{e}},
  S., Matsumura, M., Sadavoy, S., Hull, C. L.~H., Shimajiri, Y., Furuya, R.~S.,
  Johnstone, D., Plume, R., Inutsuka, S.-i., Kwon, J., \& Tamura, M. 2021,
  arXiv: 2104.11932

\bibitem[{Draine(2011)}]{Draine2011}
Draine, B.~T. 2011, {Physics of the interstellar and intergalactic medium}

\bibitem[{El-Badry {et~al.}(2019)El-Badry, Ostriker, Kim, Quataert, \&
  Weisz}]{El-Badry2019}
El-Badry, K., Ostriker, E.~C., Kim, C.-G., Quataert, E., \& Weisz, D.~R. 2019,
  MNRAS, 490, 1961

\bibitem[{Elmegreen(2011)}]{Elmegreen2011}
Elmegreen, B.~G. 2011, EAS Publ. Ser., 51, 45

\bibitem[{Fabrika {et~al.}(2021)Fabrika, Atapin, Vinokurov, \&
  Sholukhova}]{Fabrika2021}
Fabrika, S.~N., Atapin, K.~E., Vinokurov, A.~S., \& Sholukhova, O.~N. 2021,
  Astrophys. Bull., 76, 6

\bibitem[{Federrath {et~al.}(2008)Federrath, Klessen, \&
  Schmidt}]{Federrath2008}
Federrath, C., Klessen, R.~S., \& Schmidt, W. 2008, ApJ, 688, L79

\bibitem[{Field {et~al.}(1969)Field, Goldsmith, \& Habing}]{Field1969}
Field, G.~B., Goldsmith, D.~W., \& Habing, H.~J. 1969, ApJ, 155, L149

\bibitem[{Fielding {et~al.}(2018)Fielding, Quataert, \&
  Martizzi}]{Fielding2018}
Fielding, D., Quataert, E., \& Martizzi, D. 2018, MNRAS, 481, 3325

\bibitem[{Finkbeiner(2003)}]{Finkbeiner2003}
Finkbeiner, D.~P. 2003, ApJS, 146, 407

\bibitem[{Galli {et~al.}(2019)Galli, Loinard, Bouy, Sarro, Ortiz-Le{\'{o}}n,
  Dzib, Olivares, Heyer, Hernandez, Rom{\'{a}}n-Z{\'{u}}{\~{n}}iga, Kounkel, \&
  Covey}]{Galli2019}
Galli, P.~A., Loinard, L., Bouy, H., Sarro, L.~M., Ortiz-Le{\'{o}}n, G.~N.,
  Dzib, S.~A., Olivares, J., Heyer, M., Hernandez, J.,
  Rom{\'{a}}n-Z{\'{u}}{\~{n}}iga, C., Kounkel, M., \& Covey, K. 2019, arXiv,
  137, 1

\bibitem[{Gentry {et~al.}(2019)Gentry, Krumholz, Madau, \& Lupi}]{Gentry2019}
Gentry, E.~S., Krumholz, M.~R., Madau, P., \& Lupi, A. 2019, MNRAS, 483, 3647

\bibitem[{Green {et~al.}(2019)Green, Schlafly, Zucker, Speagle, \&
  Finkbeiner}]{Green2019}
Green, G.~M., Schlafly, E., Zucker, C., Speagle, J.~S., \& Finkbeiner, D. 2019,
  arXiv, 93

\bibitem[{Haid {et~al.}(2016)Haid, Walch, Naab, Seifried, Mackey, \&
  Gatto}]{Haid2016a}
Haid, S., Walch, S., Naab, T., Seifried, D., Mackey, J., \& Gatto, A. 2016,
  MNRAS, 460, 2962

\bibitem[{Hartmann {et~al.}(2002)Hartmann, Ballesteros‐Paredes, \&
  Bergin}]{Hartmann2002}
Hartmann, L., Ballesteros‐Paredes, J., \& Bergin, E.~A. 2002, ApJ, 562, 852

\bibitem[{Heitsch {et~al.}(2006)Heitsch, Slyz, Devriendt, Hartmann, \&
  Burkert}]{Heitsch2006}
Heitsch, F., Slyz, A.~D., Devriendt, J. E.~G., Hartmann, L.~W., \& Burkert, A.
  2006, ApJ, 648, 1052

\bibitem[{Hennebelle \& Inutsuka(2019)}]{Hennebelle2019}
Hennebelle, P. \& Inutsuka, S.-I. 2019, Front. Astron. Sp. Sci., 6

\bibitem[{Hennebelle \& P{\'{e}}rault(1999)}]{Hennebelle1999}
Hennebelle, P. \& P{\'{e}}rault, M. 1999, A{\&}A, 351, 309

\bibitem[{Hosokawa \& Inutsuka(2006)}]{Hosokawa2006}
Hosokawa, T. \& Inutsuka, S.-i. 2006, ApJ, 648, L131

\bibitem[{Inoue \& Inutsuka(2008)}]{Inoue2008}
Inoue, T. \& Inutsuka, S. 2008, ApJ, 687, 303

\bibitem[{Inoue \& Inutsuka(2009)}]{Inoue2009}
Inoue, T. \& Inutsuka, S.~I. 2009, ApJ, 704, 161

\bibitem[{Inutsuka {et~al.}(2015)Inutsuka, Inoue, Iwasaki, \&
  Hosokawa}]{Inutsuka2015}
Inutsuka, S.~I., Inoue, T., Iwasaki, K., \& Hosokawa, T. 2015, A{\&}A, 580, A49

\bibitem[{Kaaret {et~al.}(2017)Kaaret, Feng, \& Roberts}]{Kaaret2017}
Kaaret, P., Feng, H., \& Roberts, T.~P. 2017, ARA{\&}A, 55, 303

\bibitem[{Kerr {et~al.}(2021)Kerr, Rizzuto, Kraus, \& Offner}]{Kerr2021}
Kerr, R., Rizzuto, A.~C., Kraus, A.~L., \& Offner, S. S.~R. 2021,
  arxiv:2105.09338

\bibitem[{Kim \& Ostriker(2015)}]{Kim2015}
Kim, C.~G. \& Ostriker, E.~C. 2015, ApJ, 802, 99

\bibitem[{Kim {et~al.}(2017)Kim, Ostriker, \& Raileanu}]{kim2017b}
Kim, C.-G., Ostriker, E.~C., \& Raileanu, R. 2017, ApJ, 834, 25

\bibitem[{Koyama \& Inutsuka(2000)}]{Koyama2000}
Koyama, H. \& Inutsuka, S. 2000, ApJ, 1, 980

\bibitem[{Koyama \& Inutsuka(2002)}]{Koyama2002}
---. 2002, ApJ, 564, L97

\bibitem[{Krause {et~al.}(2014)Krause, Diehl, B{\"{o}}hringer, Freyberg, \&
  Lubos}]{Krause2014}
Krause, M., Diehl, R., B{\"{o}}hringer, H., Freyberg, M., \& Lubos, D. 2014,
  A{\&}A, 566, A94

\bibitem[{Kritsuk {et~al.}(2017)Kritsuk, Ustyugov, \& Norman}]{Kritsuk2017}
Kritsuk, A.~G., Ustyugov, S.~D., \& Norman, M.~L. 2017, New J. Phys., 19,
  065003

\bibitem[{Lallement {et~al.}(2019)Lallement, Babusiaux, Vergely, Katz, Arenou,
  Valette, Hottier, \& Capitanio}]{Lallement2019}
Lallement, R., Babusiaux, C., Vergely, J.~L., Katz, D., Arenou, F., Valette,
  B., Hottier, C., \& Capitanio, L. 2019, A{\&}A, 625, 1

\bibitem[{Lee \& Chen(2009)}]{Lee2009}
Lee, H.~T. \& Chen, W.~P. 2009, ApJ, 694, 1423

\bibitem[{Leike {et~al.}(2020)Leike, Glatzle, \& En{\ss}lin}]{Leike2020a}
Leike, R., Glatzle, M., \& En{\ss}lin, T.~A. 2020, A{\&}A, 138, 1

\bibitem[{Leike \& En{\ss}lin(2019)}]{Leike2019}
Leike, R.~H. \& En{\ss}lin, T.~A. 2019, A{\&}A, 631, 1

\bibitem[{Lim {et~al.}(2013)Lim, Min, \& Seon}]{Lim2013}
Lim, T.~H., Min, K.~W., \& Seon, K.~I. 2013, ApJ, 765, 107

\bibitem[{Limongi \& Chieffi(2006)}]{Limongi2006}
Limongi, M. \& Chieffi, A. 2006, ApJ, 647, 483

\bibitem[{Luhman(2018)}]{Luhman2018}
Luhman, K.~L. 2018, ApJ, 156, 271

\bibitem[{Mackey {et~al.}(2017)Mackey, Koposov, {Da Costa}, Belokurov, Erkal,
  Fraternali, McClure-Griffiths, \& Fraser}]{Mackey2017}
Mackey, A.~D., Koposov, S.~E., {Da Costa}, G.~S., Belokurov, V., Erkal, D.,
  Fraternali, F., McClure-Griffiths, N.~M., \& Fraser, M. 2017, MNRAS, 472,
  2975

\bibitem[{McKee \& Ostriker(1977)}]{McKee1977}
McKee, C.~F. \& Ostriker, J.~P. 1977, ApJ, 218, 148

\bibitem[{Mooley {et~al.}(2013)Mooley, Hillenbrand, Rebull, Padgett, \&
  Knapp}]{Mooley2013-lw}
Mooley, K., Hillenbrand, L., Rebull, L., Padgett, D., \& Knapp, G. 2013, The
  Astrophysical Journal, 771, 110

\bibitem[{Ntormousi {et~al.}(2011)Ntormousi, Burkert, Fierlinger, \&
  Heitsch}]{Ntormousi2011}
Ntormousi, E., Burkert, A., Fierlinger, K., \& Heitsch, F. 2011, ApJ, 731

\bibitem[{Ortiz-Le{\'{o}}n {et~al.}(2018)Ortiz-Le{\'{o}}n, Loinard, Dzib,
  Galli, Kounkel, Mioduszewski, Rodr{\'{i}}guez, Torres, Hartmann, Boden,
  Evans, Brice{\~{n}}o, \& Tobin}]{Ortiz-Leon2018}
Ortiz-Le{\'{o}}n, G.~N., Loinard, L., Dzib, S.~A., Galli, P.~A., Kounkel, M.,
  Mioduszewski, A.~J., Rodr{\'{i}}guez, L.~F., Torres, R.~M., Hartmann, L.,
  Boden, A.~F., Evans, N.~J., Brice{\~{n}}o, C., \& Tobin, J.~J. 2018, arXiv,
  865, 73

\bibitem[{Pakull \& Mirioni(2001)}]{Pakull2001}
Pakull, M.~W. \& Mirioni, L. 2001, in Symp. ‘New Visions X-ray Universe
  Xmm-newt. Chandra Era'

\bibitem[{Palou{\v{s}}(2014)}]{Palous2014a}
Palou{\v{s}}, J. 2014, Astrophys. Sp. Sci. Proc., 36, 181

\bibitem[{{Pavlidou} {et~al.}(2021){Pavlidou}, {Scholz}, \&
  {Teixeira}}]{Pavlidou2021}
{Pavlidou}, T., {Scholz}, A., \& {Teixeira}, P.~S. 2021, \mnras, 503, 3232

\bibitem[{Prantzos \& Diehl(1996)}]{Prantzos1996}
Prantzos, N. \& Diehl, R. 1996, Phys. Rep., 267, 1

\bibitem[{{Raiteri} {et~al.}(1996){Raiteri}, {Villata}, \&
  {Navarro}}]{Raiteri1996}
{Raiteri}, C.~M., {Villata}, M., \& {Navarro}, J.~F. 1996, \aap, 315, 105

\bibitem[{Sancisi(1974)}]{Sancisi1974}
Sancisi, R. 1974, in Galact. Radio Astron. Proc. from IAU Symp. no. 60 held
  Maroochydore Queensland, Aust. 3-7 Sept. 1973. Ed. by Frank J. Kerr Simon
  Christ. Simonson. Int. Astron. Union. Symp. no. 60, Dordrecht-Holland; B

\bibitem[{{Sathyaprakash} {et~al.}(2019){Sathyaprakash}, {Roberts}, {Walton},
  {Fuerst}, {Bachetti}, {Pinto}, {Alston}, {Earnshaw}, {Fabian}, {Middleton},
  \& {Soria}}]{Sathyaprakash2019}
{Sathyaprakash}, R., {Roberts}, T.~P., {Walton}, D.~J., {Fuerst}, F.,
  {Bachetti}, M., {Pinto}, C., {Alston}, W.~N., {Earnshaw}, H.~P., {Fabian},
  A.~C., {Middleton}, M.~J., \& {Soria}, R. 2019, \mnras, 488, L35

\bibitem[{Saury {et~al.}(2014)Saury, Miville-Desch{\^{e}}nes, Hennebelle,
  Audit, \& Schmidt}]{Saury2014}
Saury, E., Miville-Desch{\^{e}}nes, M.-A., Hennebelle, P., Audit, E., \&
  Schmidt, W. 2014, A{\&}A, 567, A16

\bibitem[{Seifried {et~al.}(2010)Seifried, Schmidt, \& Niemeyer}]{Seifried2010}
Seifried, D., Schmidt, W., \& Niemeyer, J.~C. 2010, A{\&}A, 526

\bibitem[{Shimajiri {et~al.}(2019)Shimajiri, Andr{\'{e}}, Palmeirim,
  Arzoumanian, Bracco, K{\"{o}}nyves, Ntormousi, \& Ladjelate}]{Shimajiri2019}
Shimajiri, Y., Andr{\'{e}}, P., Palmeirim, P., Arzoumanian, D., Bracco, A.,
  K{\"{o}}nyves, V., Ntormousi, E., \& Ladjelate, B. 2019, A{\&}A, 623

\bibitem[{Snowden {et~al.}(1997)Snowden, Egger, Freyberg, McCammon, Plucinsky,
  Sanders, Schmitt, Trumper, \& Voges}]{Snowden1997}
Snowden, S.~L., Egger, R., Freyberg, M.~J., McCammon, D., Plucinsky, P.~P.,
  Sanders, W.~T., Schmitt, J. H. M.~M., Trumper, J., \& Voges, W. 1997, ApJ,
  485, 125

\bibitem[{Su {et~al.}(2009)Su, Chen, Yang, Koo, Zhou, Jeong, \& Zhang}]{Su2009}
Su, Y., Chen, Y., Yang, J., Koo, B.-c., Zhou, X., Jeong, I.-g., \& Zhang, C.-g.
  2009, Astrophys. Journalastrop, 694, 376

\bibitem[{Ungerechts \& Thaddeus(1987)}]{Ungerechts1987}
Ungerechts, H. \& Thaddeus, P. 1987, ApJS, 63, 645

\bibitem[{V{\'{a}}zquez-Semadeni {et~al.}(2006)V{\'{a}}zquez-Semadeni, Ryu,
  Passot, Gonza, \& Gazol}]{Vazquez-Semadeni2006}
V{\'{a}}zquez-Semadeni, E., Ryu, D., Passot, T., Gonza, R.~F., \& Gazol, A.
  2006, ApJ, 1, 245

\bibitem[{Wit {et~al.}(2004)Wit, Testi, Palla, Vanzi, \& Zinnecker}]{Wit2004}
Wit, W.~J., Testi, L., Palla, F., Vanzi, L., \& Zinnecker, H. 2004, A{\&}A,
  425, 937

\bibitem[{Wolfire {et~al.}(2003)Wolfire, McKee, Hollenbach, \&
  Tielens}]{Wolfire2003}
Wolfire, M.~G., McKee, C.~F., Hollenbach, D., \& Tielens, A. G. G.~M. 2003,
  ApJ, 587, 278

\bibitem[{Zucker {et~al.}(2021)Zucker, Goodman, Alves, Bialy, Koch, Speagle,
  Foley, Finkbeiner, Leike, \& En{\ss}lin}]{Zucker2021_topology}
Zucker, C., Goodman, A.~A., Alves, J., Bialy, S., Koch, E.~W., Speagle, J.~S.,
  Foley, M., Finkbeiner, D.~P., Leike, R.~H., \& En{\ss}lin, T.~A. 2021, ApJ.
  Submitt.

\bibitem[{Zucker {et~al.}(2020)Zucker, Speagle, Schlafly, Green, Finkbeiner,
  Goodman, \& Alves}]{Zucker2020}
Zucker, C., Speagle, J.~S., Schlafly, E.~F., Green, G.~M., Finkbeiner, D.~P.,
  Goodman, A., \& Alves, J. 2020, A{\&}A, 633, 1

\end{thebibliography}
\end{document}